\documentclass[11pt,a4paper]{article}
\usepackage{graphicx}
\usepackage{dcolumn}
\newcommand{\be}{\begin{equation}}
\newcommand{\ee}{\end{equation}}
\newcommand{\bea}{\begin{eqnarray}}
\newcommand{\eea}{\end{eqnarray}}

\newcommand{\nn}{\nonumber}

\newcommand{\ep}{\epsilon}

\newcommand{\om}{\omega}

\newcommand{\ov}{\overline}

\newcommand{\op}{\overline\Pi}
\newcommand{\iop}{{\rm Im}\,\overline\Pi}
\newcommand{\od}{\overline D}  

\newcommand{\tom}{\widetilde{\om}}   

\newcommand{\vk}{\vec k}
\newcommand{\vp}{\vec p}

\newcommand{\vl}{\vec l}

\newcommand{\ks}{k \!\!\! /}

\newcommand{\ls}{l \!\!\! /}

\newcommand{\mn}{{\mu\nu}}
\newcommand{\del}{\partial}

\begin{document}

\setcounter{page}{1}

\title{Low mass enhanced probability of pion in hadronic matter due to its Landau cut contributions}
\author{Sabyasachi Ghosh}
\date{}
\maketitle
\begin{center}
\it{Instituto de Fisica Teorica, Universidade Estadual Paulista, 
Rua Dr. Bento Teobaldo Ferraz, 271, 01140-070 Sao Paulo, SP, Brazil}
\end{center}

\begin{abstract}
In the real-time thermal field theory, the pion self-energy
at finite temperature and density is evaluated where the different
mesonic and baryonic loops are considered. 
The interactions of
pion with the other mesons and baryons in the medium are governed
by the effective hadronic Lagrangian densities whose effective
strength of coupling constants have been determined from the experimental
decay widths of the mesons and baryons.
The detail branch cut structures of these different 
mesonic and baryonic loops are analyzed. 
The Landau cut contributions of different baryon and meson loops become only
relevant around the pion pole and it is completely appeared
in presence of medium. The in-medium spectral
function of pion has been plotted for different values of temperature,
baryon chemical potential as well as three momentum of the pion.
A noticeable low mass probability in pion spectral function promise
to contribute in the low mass dilepton enhancement via indirect modification
of $\rho$ self-energy for $\pi\pi$ loop.
\end{abstract}


\maketitle

\section{Introduction}
The recent reviews~\cite{Rapp_rev,MGM} claim that 
the calculations based on hadronic many-body theory
are very successful to describe the low mass dimuon
enhancement measured by the NA60 collaboration~\cite{NA60}, where
a strong broadening of rho meson are essentially concluded
as a main reason.
There are two sources which are generally believed to be
responsible for this broadening
of the $\rho$ meson. 
One is coming from the direct interactions of $\rho$ with the other thermalized
mesonic~\cite{Rapp_Gale,GSM_EPJC} and baryonic~\cite{Friman,Leupold,GS_NPA} 
constituents of the medium. Another is originating from the indirect interaction
of $\rho$ with medium via pion cloud, which may be modified through 
interactions with the surrounding medium~\cite{Chanfray,Herrman,Rapp_picloud}.
The latter source indicates that the in-medium modification of pion
may create an effect on $\rho$ meson propagation in the medium, for which
the rate of low mass dileptons are expected to be enhanced.
In this context, the present manuscript is intended to investigate
the in-medium modification of pion propagation due to strong interaction
with the other thermalized mesons and baryons in the medium.
For this purpose the in-medium pion self-energy for different mesonic
and baryonic loops are calculated in the real-time thermal field theory (RTF)
and their detailed branch cut structures are explicitly analyzed.
With respect to earlier 
investigations~\cite{Schenk,Chanfray_D,Shuryak,Rapp_M,Rapp_B,Rapp_B2,Tolos}, one 
of the major contribution
of this present article
is that an extensive set of baryonic loops are taken to estimate
the in-medium self-energy of pion. The $\pi\sigma$ and $\pi\rho$
loops have been taken as mesonic loops because the resonances $\sigma$
and $\rho$ have to be considered in effective hadronic model
to reproduced the experimental data of $\pi\pi$ scattering 
cross section~\cite{SSS}.

The manuscript is organized as follows.
The expression of thermal propagator for $\pi$ meson
is explicitly derived in the next section and then the detail calculations of pion
self-energy at finite temperature and density are addressed in 
Sec. (3). The baryon and meson loop calculations are separately
discussed in two subsection of the Sec. (3).
The detailed numerical results are discussed in Sec. (4)
and at last section the intention of the article is summarized.

\section{Pion propagator in the medium}
In the real-time formulation of thermal field theory,
any two point function like propagator acquires $2\times 2$ matrix structure.
The free propagator matrix of pion at finite temperature can be defined
as
\be
D_{ab}(k)=i\int d^4x e^{ikx}\langle T_c\pi(x)\pi(0) \rangle_{ab}
\ee
where the subscripts $a, b(= 1, 2)$ are thermal
indices, $\pi(x)$ is pion field and $T_c$ denotes time ordering with respect to a symmetrical 
contour in the plane of the complex time variable~\cite{sym}.
The notation $\langle {\cal O} \rangle$ is used to represent
ensemble average of any operator ${\cal O}$, 
$\langle {\cal O} \rangle={\rm Tr}\frac{e^{\beta H}{\cal O}}{{\rm Tr}e^{\beta H}}$,
where Tr indicates trace over a complete set of states.
With the help of the Dyson
equation, the complete in-medium $\pi$-propagator can be expressed as 
\be
D_{ab}(k)=D_{ab}^{(0)}(k)- D_{ac}^{(0)}(k)\Pi_{cd}(k) D_{db}(k)
\label{dyson_G_vac}
\ee
where $D_{ab}^{(0)}(k)$ and $\Pi_{ab}(k)$ are free propagator and self-energy
matrices of pion at finite temperature. 
By the diagonalization procedure~\cite{Kobes} the thermal indices ($a,b,c,d$)
can be removed and it gives the following equation :
\be
\od(k)=\od^{(0)}(k)-\od^{(0)}(k)\op(k)
\od(k)
\label{dyson_G_diag}
\ee
where the quantities with bars represent the corresponding diagonal components.
The diagonal element of free propagator is expressed as
\be
\od^{(0)}(k)=\frac{-1}{k^2-m_\pi^2+i\eta}
\label{vac}
\ee
which is exactly similar with the vacuum
free propagator.
Using Eq.~(\ref{vac}) in (\ref{dyson_G_diag}) we get the diagonal element
of complete pion propagator, 
\be
\od(k)=\frac{-1}{k^2-m_\pi^2-\op(k)}
\label{D_comp}
\ee
whose imaginary part provides the Breit-Wigner type structure
of pion spectral function,
\be
A_\pi(k)={\rm Im}\od(k)=\frac{-{\rm Im}\op(k)}{\{k^2-m_\pi^2-{\rm Re}\op(k)\}^2
+\{{\rm Im}\op(k)\}^2}~.
\label{pi_spec}
\ee
Here the information of medium is embedded in the imaginary and real part of
the pion self-energy (${\rm Im}\op$ and ${\rm Re}\op$) which will
be derived in next subsection.
\begin{figure}
\begin{center}
\includegraphics[scale=1.0]{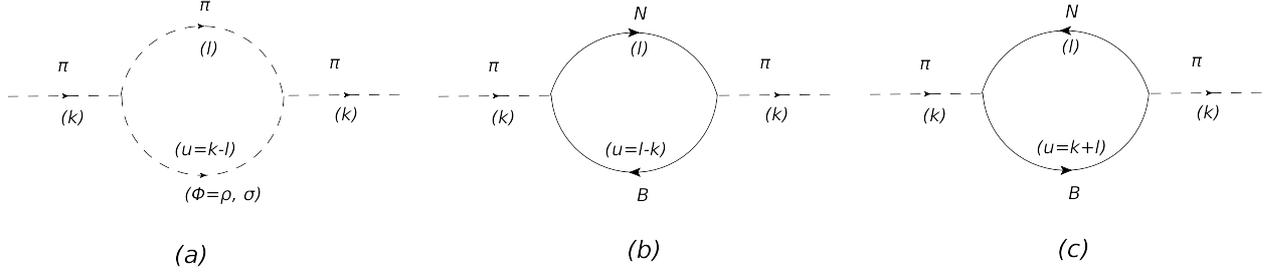}
\end{center}
\caption{Self-energy diagrams of pion for meson loops ($\pi\Phi$)
is shown in diagram (a). The baryon loops ($NB$) of pion
self-energy can be represented in two possible ways - (b) and (c).}
\label{pion_self}
\end{figure} 

\section{Pion self-energy in the medium}
In this section, the different meson and baryon loop diagrams
(shown in Fig.~\ref{pion_self}) of $\pi$ meson,  have been evaluated 
in order to investigate the in-medium modification of pion in hadronic
matter. The calculations of meson and baryon loops 
are separately addressed
below.
\subsection{Meson loops}
From the lowest order $\chi$PT, the estimated $\pi\pi$ 
cross section in free space is well in agreement with the experimental data 
up to the center-of-mass energy $\sqrt{s}= 0.5$ GeV. Beyond this value of 
$\sqrt{s}$, the $\sigma$ and $\rho$ resonances play an important role to 
explain the experimental data. By the unitarization technique, 
the $\sigma$ and $\rho$ resonances can be 
generated dynamically~\cite{Nicola} in the amplitude. An alternative 
way~\cite{SSS,GKS_PRC}, which is followed in the present paper, 
is to incorporate these resonances 
by using the effective Lagrangian densities for $\pi\pi\sigma$ and $\pi\pi\rho$ 
interactions: 
\be
{\cal L} = g_\rho \, {\vec \rho}_\mu \cdot {\vec \pi} \times \del^\mu {\vec \pi} 
+ \frac{g_\sigma}{2} m_\sigma {\vec \pi}\cdot {\vec\pi}\,\sigma,
\label{Lag}
\ee
where the coupling constants ($g_\sigma=5.82$ and $g_\rho=6$) are fixed from
their experimental decay widths of $\sigma$ and $\rho$ mesons
in their $\pi\pi$ channels~\cite{GKS_PRC}.
Serious defect of phenomenological interactions is that 
they often do not respect constraints from chiral symmetry. 
Owing to the appearance of $\sigma$ and $\rho$
resonances in $\pi\pi$ scattering cross section,
we have considered $\pi\sigma$ and $\pi\rho$ loops
to calculate in-medium pion self-energy where the interaction
part is governed by the effective Lagrangian densities, mentioned
in Eq.~(\ref{Lag}).
This pion self-energy function for $\pi\sigma$ and $\pi\rho$ loops 
has diagrammatically been represented
by the Fig.\ref{pion_self}(a).
The 11-component of pion self energy for the 
$\pi \Phi$ loop (where $\Phi$ denotes $\sigma$ or $\rho$ in generalized way) 
is given by
\be
\Pi^{11}_{M}(k)=i 
\int \frac{d^4l}{(2\pi)^4}L(k,l) D^{11}(l,m_{\pi})D^{11}(k-l, m_{\Phi})
\label{self_meso}
\ee
where the vertex factors,
\be
L(k,l) = - \frac{g^2_\sigma m_\sigma^2}{4}, 
\ee
for the $\pi\sigma$ loop, and 
\be
L(k,l) = -\frac{g^2_\rho}{m_\rho^2} \, 
\Bigl\{ k^2 \left(k^2 - m^2_\rho\right) + 
l^2 \left(l^2 - m^2_\rho\right) - \, 2\left[ 
\left(k\cdot l\right) \, m^2_\rho + k^2 \,l^2 \right]
\Bigr\},
\ee
for the $\pi\rho$ loop have been obtained from the effective 
hadronic Lagrangian, given in Eq.~(\ref{Lag}). 
The $D^{11}(l,m_\pi)$ and $D^{11}(k-l, m_{\Phi})$ in (\ref{self_meso}) 
are the 11-component 
of the scalar propagator for $\pi$ and $\Phi$ meson respectively. For example,
\be
D^{11}(l,m_\pi)=\frac{-1}{l^2-m_\pi^2+i\eta}+2i\pi n_l\delta(l^2-m_\pi^2)
\ee
where $n_l=\frac{1}{e^{\beta\om_l}-1}$ is Bose-Einstein distribution 
function of pion with its on-shell energy, 
$\om_l=\sqrt{\vl^2+m_\pi^2}$.
Similar to the propagator matrix, the self-energy
matrix can also be diagonalized into a single component
and the diagonal element and
11 component are related as~\cite{Kobes,Ghosh_thesis}
\bea
{\rm Im}\op_M(k)&=&{\rm tanh}\left(\frac{\beta k_0}{2}
\right){\rm Im}\Pi^{11}_M(k)
\nn\\
{\rm Re}\op_M(k)&=&{\rm Re}\Pi^{11}_M(k)~.
\label{R_bar_N}
\eea

After doing the $l_0$ integration of Eq.~(\ref{self_meso}) 
and then using the Eq.~(\ref{R_bar_N}),
the diagonal element of the pion self-energy matrix 
is obtained as~\cite{Ghosh_thesis}
\bea
{\ov\Pi}_{M}(k)&=&\int\frac{d^3l}{(2\pi)^3}\frac{1}
{4\om_{l}\om_{u}}\left[\frac{(1+n_l)L_1+n_{u}L_3}
{k_0 -\om_{l}-\om_{u}+i\eta\ep(k_0)}
+\frac{-n_{l}L_1+n_{u}L_4}
{k_0-\om_{l}+\om_{u}+i\eta\ep(k_0)} 
\right.\nn\\
&&+\left. \frac{n_{l}L_2 -n_{u}L_3}{k_0 +\om_{l}-\om_{u}+i\eta\ep(k_0)} 
+\frac{-n_{l}L_2 -(1+n_{u})L_4}
{k_0 +\om_{l}+\om_{u}+i\eta\ep(k_0)}\right]
\label{MM_rho}
\eea
where $L_i,i=1,..4$ denote the values of $L(l_0)$ for
$l_0=\om_l,-\om_l,k_0-\om_u,k_0+\om_u$ respectively
and $n_u=\frac{1}{e^{\beta\om_u}-1}$ is Bose-Einstein distribution 
function for $\Phi$ resonance with its on-shell energy, 
$\om_u=\sqrt{(\vk-\vl)^2+m_\Phi^2}$. 
The imaginary part in the relevant Landau and unitary cut
regions are respectively given below
\be
{\rm Im}{\ov\Pi}_{M}=-\frac{\ep(q_0)}{16\pi |\vk|}\int_{{\tom^+_{l}}}^{{\tom^-_{l}}} d\tom_{l} L_2
\{n(\tom_{l})-n(\tom_{u}=k_0+\tom_{l})\}
\label{im_L}
\ee
and
\be
{\rm Im}{\ov\Pi}_{M}=-\frac{\ep(q_0)}{16\pi |\vk|}\int_{\om^-_{l}}^{\om^+_{l}} d\om_l L_1
\{1+n(\om_{l})+n(k_0-\om_{l})\}
\label{im_U}
\ee
where the integration limits
$\om^{\pm}_{l}=\frac{S^2_{\pi}}{2k^2}(k_0\pm |\vk|W_{\pi})$, 
${\tom^{\pm}_{l}}=\frac{S^2_{\pi}}{2k^2}(-k_0\pm |\vk|W_{\pi})$
with $W_{\pi}=\sqrt{1 -\frac{4k^2 m_\pi^2}{S^4_{\pi}}}$ 
and $S^2_{\pi}=k^2-m^2_{\Phi}+m^2_{\pi}$. These two regions
of branch cuts are $\vk<k_0<\sqrt{\vk^2+(m_\Phi-m_\pi)^2}$
and $\sqrt{\vk^2+(m_\Phi+m_\pi)^2}<k_0<\infty$ respectively.

The real part of the self-energy can be easily read off from (\ref{MM_rho}) 
in terms of principal value integrals and we do not write them here.

In realistic scenario, the $\sigma$ meson should not be treated
as stable particle as considered in our previous framework
as it exhibits a broad spectral function in vacuum.
Hence, to take into account its broad width, 
the Eq.~(\ref{MM_rho}) has been convoluted as (see e.g.~\cite{GKS_PRC,Oset,GS_NPA,S_omega})
\be
{\ov\Pi}_{M}(k, m_\sigma) = \left[\int^{(m^+_\sigma)^2}_{(m^-_\sigma)^2}
dp^2  \, A_\sigma(p) \, {\ov\Pi}_{M}(k, p)\right] 
/ \left[\int^{(m^+_\sigma)^2}_{(m^-_\sigma)^2}
dp^2\; A_\sigma(p)\right] ,
\label{gm_mu}
\ee
where ${\ov\Pi}_{M}(k, p)$ is the narrow-width expression self-energy 
given in Eq.~(\ref{MM_rho}), with $m_\sigma$ replaced by $p$; 
$A_\sigma(p^2)$ is the spectral density:
\be
A_\sigma(p^2) = \frac{1}{\pi}{\rm Im}\left[\frac{-1}{p^2-m_\sigma^2+ip\Gamma_\sigma(p)}\right]~.
\label{rho_u}
\ee
$\Gamma_\sigma(p)$ is the vacuum spectral widths of the mesons:
\be
\Gamma_\sigma (p) = \frac{3g_\sigma^2m_\sigma^2}{32\pi p} 
\left(1 - \frac{4m_\pi^2}{p^2}\right)^{1/2}~.
\label{s-width}
\ee
In the integration limits, $m^\pm_\sigma = m_\sigma \pm 2\,\Gamma^0_\sigma$, 
with $\Gamma^0_\sigma = \Gamma_\sigma(p=m_\sigma)$. 

\subsection{Baryon loops} 
Next we will calculate the baryon loops of pion
self-energy in the same framework of RTF.
Besides the meson loops in the medium,
the pion propagator may also be undergone via
different intermediate $NB$ loops, where $B$ stands for different
higher mass baryons including nucleon itself. In this work
we have taken
an extensive set of 4-star baryon resonances with spin one-half and three-half.
These are $N(940)$, $\Delta(1232)$, $N^*(1440)$, $N^*(1520)$,
$N^*(1535)$, $\Delta^*(1600)$, $\Delta^*(1620)$, $N^*(1650)$, 
$\Delta^*(1700)$, $N^*(1700)$, $N^*(1710)$, $N^*(1720)$
where their mass (in MeV) are presented inside the brackets.
The direct and cross diagrams of pion self-energy 
for $NB$ loops have been represented
in the diagram~\ref{pion_self}(b) and (c).

The $11$-component of the in-medium pion self-energy for the $NB$ loop 
is given by
\be
\Pi^{11}_B(k)=i \sum_{a=-1,+1}
\int \frac{d^4l}{(2\pi)^4}  L(k,l)
E^{11}(l,m_N) E^{11}(l-a k,m_B) 
\label{Pi11B}
\ee
where $E^{11}(l,m_N)$ and $E^{11}(l-ak,m_B)$ are
respectively
scalar part of the nucleon and 
baryon propagators at finite temperature. 
In RTF its expression is as follows : 
\bea
E^{11}(l,m_N)&=&\frac{-1}{l^2-m_N^2+i\eta}-2\pi i
N(l_0)\delta(l^2-m_N^2);~~~~~~~~N(l_0)=n^+_l\theta(l_0)+
n^-_l\theta(-l_0)\nonumber\\
\nonumber\\
&=&-\frac{1}{2\om_l}\left(\frac{1-n^+_l}{l_0-\om_l+i\eta}+
\frac{n^+_l}{l_0-\om_l-i\eta}-\frac{1-n^-_l}{l_0+\om_l-i\eta}
-\frac{n^-_l}{l_0+\om_l+i\eta}\right)
\label{de11}
\eea
where $n^{\pm}_l=\displaystyle\frac{1}{e^{\beta(\om_l \mp \mu_N)}+1}$
is the Fermi-Dirac distribution for energy $\om_l=\sqrt{\vl^2+m_N^2}$ and the $\pm$ 
sign in the superscript of $n_l$ refer to nucleon and 
anti-nucleon respectively.  
Here $\mu_N$ is the chemical potential of nucleon which is 
supposed to be equal with the chemical potentials
of all the baryons considered here. 
The two values of $a$ in (\ref{Pi11B}) correspond to the direct 
and crossed diagrams shown in Fig.~\ref{pion_self}
(b) and (c) respectively which can be obtainable from one another by changing the 
sign of the external momentum $k$.

Let us first discuss diagram (b) for which $a=+1$.
After integrating over $l^0$ in Eq.~(\ref{Pi11B})
and then using the similar kind of relation like Eq.~(\ref{R_bar_N}),
the diagonal element of the in-medium self energy can be expressed as
\bea
{\ov\Pi}_B(k)&=&\int\frac{d^3l}{(2\pi)^3}\frac{1}{4\om_l\om_u}
\left[\frac{(1-n_l^+)L_1-n_u^-L_3}{k_0 -\om_l-\om_u+i\eta\ep(k_0)}
+\frac{n_l^+L_1-n_u^+L_4}{k_0-\om_l+\om_u+i\eta\ep(k_0)} 
\right.\nn\\
&&+ \left.\frac{-n_l^-L_2 +n_u^-L_3}{k_0 +\om_l-\om_u+i\eta\ep(k_0)} 
+\frac{n_l^-L_2 +(-1+n_u^+)L_4}{k_0 +\om_l+\om_u+i\eta\ep(k_0)}\right]
\label{Pi_a}
\eea
where $n^{\pm}_u$ are also Fermi-Dirac distribution functions
for baryon and anti-baryon with $\om_u=\sqrt{(\vl-\vk)^2+m_{B}^2}$
and $L_i,i=1,..4$ denote the values of $L(l_0)$ for
$l_0=\om_l,-\om_l,k_0-\om_u,k_0+\om_u$ respectively. 
For the diagram (c) in Fig.~\ref{pion_self}, a similar kind
of expression like Eq.~(\ref{Pi_a}) can also be received just by
putting $a=-1$.
The imaginary part of the third term of (\ref{Pi_a}) will be
non-zero in the Landau region $\vk<k_0<\sqrt{\vk^2+(m_B-m_N)^2}$
and only this contribution is significant for pion spectral function
because pion pole is situated in this region while unitary
cuts are far from it.
Adding the relevant Landau cut contributions of diagram (b) and (c),
the total contribution from the baryon
loops is given by
\bea
\iop_B(k_0,\vk)&=&\frac{-\ep(k_0)}{16\pi|\vk|}\int_{\tom^{+}_{l}}^{\tom^{-}_{l}} 
d\tom_l [L_1(a=-1)\{-n_+(\tom_l)+n_+(\tom_u=k_0+\tom_l)\}
\nn\\
&&~~~~~~~~~~~~~~+L_2(a=+1)\{-n_-(\tom_l)+n_-(\tom_u=k_0+\tom_l)\}]
\label{Pi_B_bar}
\eea
where $\tom^{\pm}_{l}=\frac{S^2_{N}}{2k^2}(-k^0 \pm |\vk| W_{N})$
with $W_{N}=\sqrt{1-\frac{4k^2m_N^2}{S^4_N}}$,
$S^2_{N}=k^2-m_B^2+m_N^2$.
For the $B=N$ (i.e. $NN$ loop), the diagram (b) and (c) are identical, hence
only one of them has to be included.

Similar to mesonic case, the real part of $\iop_B$ can also be
extracted from Eq.~(\ref{Pi_B_bar}) by taking its principal
value integral. The thermal part of Re$\Pi_B$ or Re$\Pi_M$
is the main interest to estimate the mass shift of pion,
which can be reflected from the shifting of peak position
in the pion spectral function.

\begin{table}[h]
\begin{center}
\begin{tabular}{|c|c|c|c|c|c|}
\hline
& & & & & \\
Baryons & $J_B^P$ & $I_B$ & $\Gamma_{\rm tot}$ & $\Gamma_{B\rightarrow N\pi}$ (B.R.) & $f/m_\pi$ \\
& & & & &  \\
\hline
& & & & & \\
$\Delta(1232)$ & ${\frac{3}{2}}^+$ & 3/2 & 0.117 & 0.117 (100\%) & 15.7 \\
& & & & & \\
$N^*(1440)$ & ${\frac{1}{2}}^+$ & 1/2 & 0.300 & 0.195 (65\%) & 2.5 \\
& & & & & \\
$N^*(1520)$ & ${\frac{3}{2}}^-$ & 1/2 & 0.115 & 0.069 (60\%) & 11.6 \\
& & & & & \\
$N^*(1535)$ & ${\frac{1}{2}}^-$ & 1/2 & 0.150 & 0.068 (45\%) & 1.14 \\
& & & & & \\
$\Delta^*(1600)$ & ${\frac{3}{2}}^+$ & 3/2 & 0.320 & 0.054 (17\%) & 3.4 \\
& & & & & \\
$\Delta^*(1620)$ & ${\frac{1}{2}}^-$ & 3/2 & 0.140 & 0.035 (25\%) & 1.22 \\
& & & & & \\
$N^*(1650)$ & ${\frac{1}{2}}^-$ & 1/2 & 0.150 & 0.105 (70\%) & 1.14 \\
& & & & & \\
$\Delta^*(1700)$ & ${\frac{3}{2}}^-$ & 3/2 & 0.300 & 0.045 (15\%) & 9.5 \\
& & & & & \\
$N^*(1700)$ & ${\frac{3}{2}}^-$ & 1/2 & 0.100 & 0.012 (12\%) & 2.8 \\
& & & & & \\
$N^*(1710)$ & ${\frac{1}{2}}^+$ & 1/2 & 0.100 & 0.012 (12\%) & 0.35 \\
& & & & & \\
$N^*(1720)$ & ${\frac{3}{2}}^+$ & 1/2 & 0.250 & 0.028 (11\%) & 1.18 \\
& & & & & \\
\hline
\end{tabular}
\label{Tab1}
\caption{From the left to right columns, the table contain
the baryons, their spin-parity quantum numbers $J_B^P$,
isospin $I_B$, total decay width $\Gamma_{\rm tot}$,
decay width in $N\pi$ channels $\Gamma_{B\rightarrow N\pi}$ or
$\Gamma_B$ in Eq.~(\ref{Gam_BNpi}) (brackets displaying 
its Branching Ratio) and at the last coupling constants $f/m_\pi$.}
\end{center}
\end{table}

\begin{figure}
\begin{center}
\includegraphics[scale=0.35]{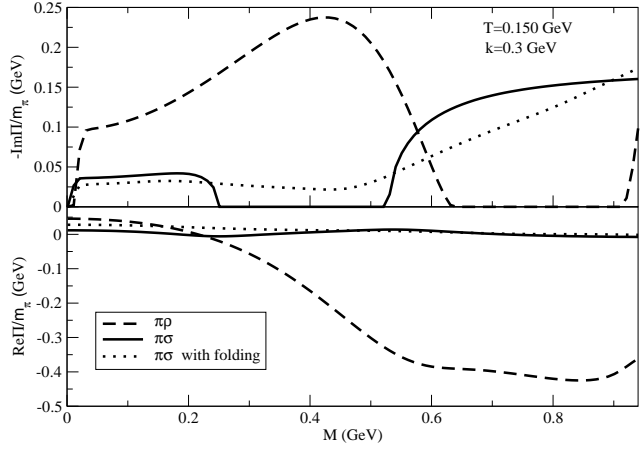}
\end{center}
\caption{Imaginary (upper panel) and real (lower panel)
part of pion self-energy  for $\pi\sigma$ (solid line) 
and $\pi\rho$ (dashed line) loops.
In the imaginary part of the pion self-energy, two separate regions 
of Landau and unitary cuts are distinctly revealed. The dotted
line represents the $\pi\sigma$ loop contribution after foding
by the broad vacuum spectral function of $\sigma$.}
\label{pi_mes}
\end{figure}
To calculate $L(k,l)$ for baryon loops we have used
the effective $BN\pi$ interaction Lagrangian densities~\cite{Leupold_pi}
\bea
{\cal L}&=&\frac{f}{m_\pi}{\ov \psi}_B\gamma^\mu
\left\{
\begin{array}{c}
i\gamma^5 \\
1
\end{array}
\right\}
\psi_N\del_\mu\pi + {\rm h.c.}~{\rm for}~J_B^P=\frac{1}{2}^{\pm}
\nn\\
&=&\frac{f}{m_\pi}{\ov \psi}^\mu_B
\left\{
\begin{array}{c}
1 \\
i\gamma^5
\end{array}
\right\}
\psi_N\del_\mu\pi + {\rm h.c.}~{\rm for}~J_B^P=\frac{3}{2}^{\pm}~,
\label{Lag_BNpi}
\eea
where $J_B$ and $P$ are spin and parity quantum numbers of 
corresponding baryon, $B$.
The effective strength of the coupling constants $f/m_\pi$ 
for different $BN\pi$ interactions can be determined from
the experimental vacuum widths of corresponding $B\rightarrow N\pi$
decays.
From above Lagrangian densities, one can easily
deduce
\bea
L(k,l)&=&-\left(\frac{f}{m_\pi}\right)^2{\rm Tr}[\ks(\ls-a\ks -Pm_B)
\ks(\ls +m_N)]
\nn\\
&=&-4\left(\frac{f}{m_\pi}\right)^2[2(k\cdot l)^2-a(k\cdot l)k^2-k^2(l^2+m_Nm_B)]
\eea
for $J_B^P=\frac{1}{2}^{\pm}$ and
\bea
L(k,l)&=&-\left(\frac{f}{m_\pi}\right)^2{\rm Tr}
\left[(\ls +m_N)(\ls-a\ks +Pm_B)k_\mu k_\nu
\left\{-g^{\mn}+\frac{1}{3}\gamma^\mu\gamma^\nu
\right.\right.\nn\\
&&\left.\left.+\frac{2}{3m_B^2}(l-ak)^\mu(l-ak)^\nu
+\frac{1}{3m_B}(\gamma^\mu(l-ak)^\nu-(l-ak)^\mu\gamma^\nu)\right\}\right]
\nn\\
&=&-\frac{8}{3m_B^2}\left(\frac{f}{m_\pi}\right)^2[m_Nm_B+l^2-a(k\cdot l)]
[(l\cdot k-ak^2)^2-k^2m_B^2]
\eea
for $J_B^P=\frac{3}{2}^{\pm}$. 
Following the Ref.~\cite{Rapp_rev}, a soft monopole form factor
$F=(\Lambda^2-m_\pi^2)/(\Lambda^2+\vk^2)$ with $\Lambda=0.3$ GeV has been multiplied
in the $\pi NB$ vertex to take into account its finite size.

The Lagrangian densities in (\ref{Lag_BNpi}) are not displaying
its isospin structures. 
For 
$J_B^P={\frac{1}{2}}^\pm$ and $J_B^P={\frac{3}{2}}^\pm$,
these isospin structures should be ${\ov \psi}{\vec\tau}\cdot{\vec\pi}\psi$ and
${\ov \psi}{\vec T}\cdot{\vec\pi}\psi$ respectively, where
${\vec T}$ and ${\vec \tau}$ stand for the usual spin $3/2$ transition
and Pauli operator. These isospin structures provide 
appropriate isospin factors, which have to be multiplied with the expressions of 
corresponding $N B$ loop diagrams. 
The isospin factor for $N N$ or $N N^*$ loops is
$I_{\pi\rightarrow N N,N^*}=2$ and for the $N\Delta$
or $N\Delta^*$, it is $I_{\pi\rightarrow N \Delta,\Delta^*}=4/3$. 

Next we have calculated vacuum width of different baryons in the $N\pi$
decay channel to fix their corresponding coupling constants $f/m_\pi$. 
With the help of the Lagrangian densities (\ref{Lag}), vacuum
decay width of baryons $B$ for $N\pi$ channel can be
obtained as
\bea
\Gamma_B&=&\frac{I_{N^*\rightarrow\pi N}}{2J_B+1}
\left(\frac{f}{m_\pi}\right)^2\frac{|\vp_{cm}|}{2\pi m_B}
[2m_B|\vp_{cm}|^2
+m_\pi^2(\om_N-Pm_N)]~~~{\rm for}~J_B^P=\frac{1}{2}^{\pm}
\nn\\
&=&\frac{I_{\Delta,\Delta^*\rightarrow\pi N}}{2J_B+1}
\left(\frac{f}{m_\pi}\right)^2\frac{|\vp_{cm}|^3}{3\pi m_B}
[\om_N+Pm_N]~{\rm for}~~~J_B^P=\frac{3}{2}^{\pm}
\label{Gam_BNpi}
\eea
where $|\vp_{cm}|=\frac{\sqrt{\{m_B^2-(m_N+m_\pi)^2\}\{m_B^2-(m_N-m_\pi)^2\}}}{2m_B}$
and $\om_N=\sqrt{|\vp_{cm}|^2+m_N^2}$.
The isospin factors are
$I_{N^*\rightarrow\pi N}=3$ and $I_{\Delta,\Delta^*\rightarrow\pi N}=1$
for the $N\pi$ decay channels of $N^*$ and $\Delta^*$ (or $\Delta$)
respectively.
Putting the experimental values~\cite{PDG} of $\Gamma_B$ in Eq.~(\ref{Gam_BNpi}),
the values of coupling constants $f/m_\pi$ have been fixed, which are 
shown in a Table~(1).

\section{Results and discussion}
\begin{figure}
\begin{center}
\includegraphics[scale=0.35]{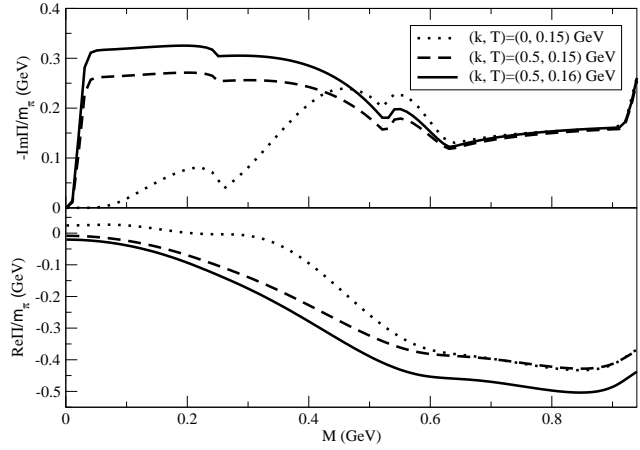}
\end{center}
\caption{Imaginary (upper panel) and real (lower panel)
part of total pion self-energy coming from the meson loops
are presented for different set of temperature $T$
and three momentum $\vk$.}
\label{pi_mes_kT}
\end{figure}
\begin{figure}
\begin{center}
\includegraphics[scale=0.35]{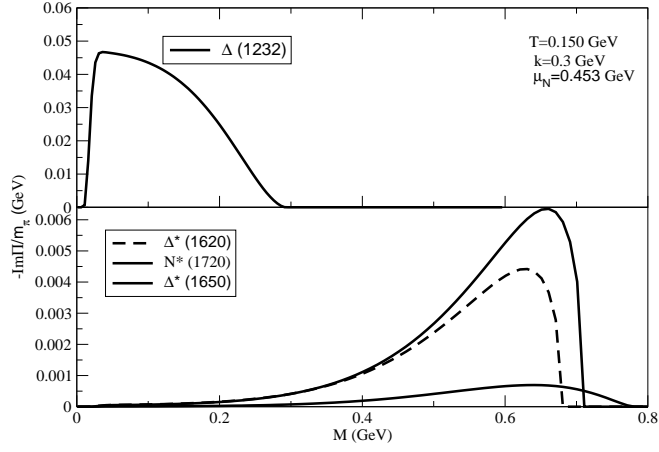}
\end{center}
\caption{Imaginary part of pion self-energy for different $NB$ 
loops. $B= \Delta^*(1620),~ N^*(1650),~N^*(1720)$ are displayed in 
lower panel while $B=\Delta(1232)$  
is shown in upper panel. we have taken the medium parameters $T=0.15$ GeV,
$\mu_N=0.453$ GeV, for which the nucleon density become $\rho_N=0.55\rho_0$ ($\rho_0$
is saturation nuclear matter density).}
\label{piIm_M}
\end{figure} 
\begin{figure}
\begin{center}
\includegraphics[scale=0.35]{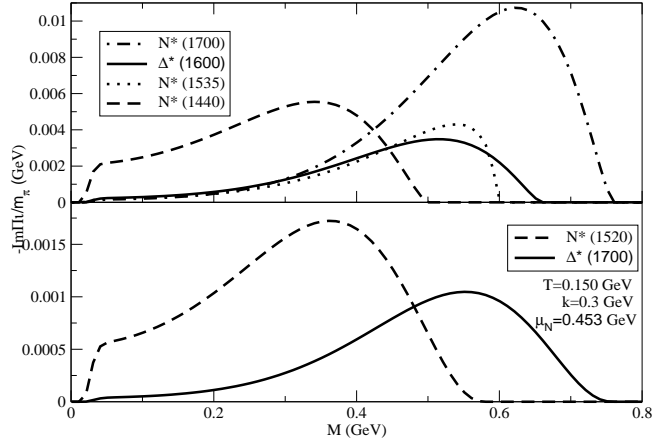}
\end{center}
\caption{Same as Fig.~(\ref{piIm_M}) for rest of the baryons
$B=N^*(1440),~ N^*(1535),~\Delta^*(1600)~N^*(1700)$ (upper panel) and
$B=N^*(1520),~\Delta^*(1700)$ (lower panel).}
\label{piIm_M2}
\end{figure} 
\begin{figure}
\begin{center}
\includegraphics[scale=0.35]{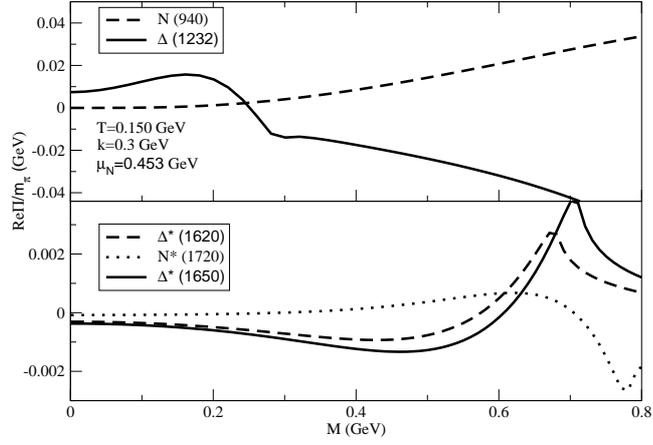}
\end{center}
\caption{The corresponding
results of Fig.~(\ref{piIm_M}) for the real part
of pion self-energy. Only $NN$ loop contribution is
additionally presented here which has vanishing Landau cut
contribution in Fig.~(\ref{piIm_M}).}
\label{piRe_M}
\end{figure} 
\begin{figure}
\begin{center}
\includegraphics[scale=0.35]{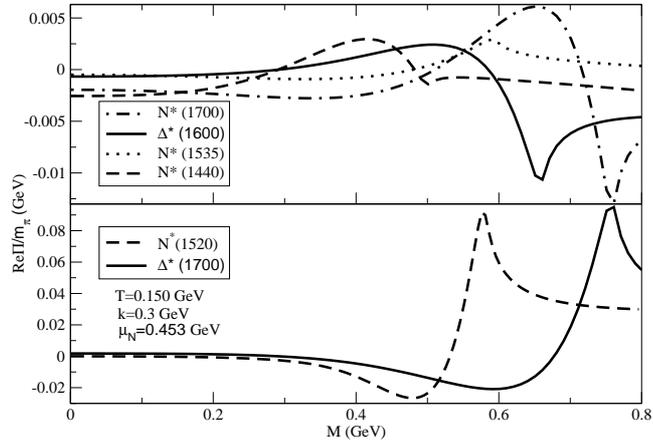}
\end{center}
\caption{The corresponding
results of Fig.~(\ref{piIm_M2}) for the real part
of pion self-energy..}
\label{piRe_M2}
\end{figure} 
Now let us discuss the results of numerical evaluation.
In Fig.~(\ref{pi_mes}), the imaginary (upper panel) and real (lower panel) 
part of pion self-energy for $\pi\sigma$ (solid line) and $\pi\rho$ (dashed line) 
loops are plotted against the invariant mass
$M$, where $M=\sqrt{k_0^2-\vk^2}$. The Landau and unitary regions
for $\pi\sigma$ and $\pi\rho$ loops are distinctly observed
from the upper panel of the Fig.~(\ref{pi_mes}).
The Landau regions of the $\pi\sigma$ and $\pi\rho$
loops are respectively ($M=0$ to $m_\sigma-m_\pi$ i.e.
$0$ to $0.25$ GeV) and ($M=0$ to $m_\rho-m_\pi$ i.e.
$0$ to $0.63$ GeV) whereas their unitary regions start respectively
from ($M=m_\sigma+m_\pi=0.53$ GeV) and ($M=m_\rho+m_\pi=0.91$ GeV) 
to $\infty$. 
Since the unitary regions are far away from the pion pole ($M=m_\pi=0.14$ GeV),
the unitary cut contributions may not play any dominating role in
the spectral profile of pion.
Rather the Landau cut contributions seem to be responsible
for determining the spectral profile of pion.
For $\pi\sigma$ loop, the broad vacuum spectral shape of $\sigma$
is taken into account by using Eq.~(\ref{gm_mu}) and the corresponding
results are shown by dotted line in Fig.~(\ref{pi_mes}). From the upper
panel of the Fig.~(\ref{pi_mes}), the dotted line indicates that
due to the folding, the Landau
and unitary cut regions are overlaped by each other and their contributions
are also reduced. 
For the different set of temperature $T$ and the three momentum $\vk$,
the imaginary (upper panel) and real (lower panel) part of
total self-energy after summing the contributions of $\pi\sigma$
and $\pi\rho$ loops have been shown in the Fig.~(\ref{pi_mes_kT}).
Here we see that the numerical strength of imaginary part of pion
self-energy is significantly enhanced in the low $M$ regions (including
the pion pole) for finite $\vk$. It also increases with the increasing 
of the $\vk$ and $T$. The real part of the pion self-energy become very 
small around it's pole position. 

\begin{figure}
\begin{center}
\includegraphics[scale=0.35]{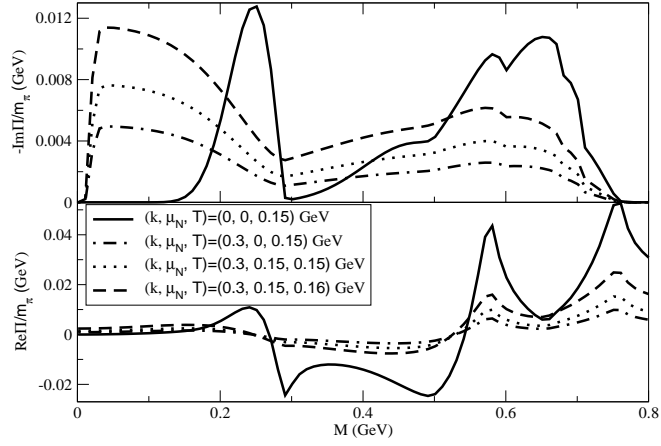}
\end{center}
\caption{Imaginary (upper panel) and real (lower panel)
part of total pion self-energy coming from the baryon loops
are presented for different set of temperature $T$, baryon
chemical potential $\mu_N$ and three momentum $\vk$.}
\label{pi_B_kT}
\end{figure}

Next we will discuss about the results of pion self-energy
which are coming from the different $NB$ loops.
The Fig.~(\ref{piIm_M}) demonstrates the imaginary part of pion
self-energy for the baryons
$B=\Delta(1232)$ (upper panel)
and $B=\Delta^*(1620)$, $N^*(1650)$, $N^*(1720)$ (lower panel).
Whereas the results for baryons 
$B=N^*(1440)$, $N^*(1535)$, $\Delta^*(1600)$, $N^*(1700)$ (upper panel)
and $B=N^*(1520), \Delta^*(1700)$ (lower panel) are displayed
in the Fig.~(\ref{piIm_M2}). The result of the 
$B=N^*(1710)$ is  not revealed with the other baryons as its strength is very low.
In those figures the Landau regions of different loops are clearly identified.
As an example 
the Landau region of the $N\Delta$ loop is ($M=0$ to $m_\Delta-m_N$ i.e.
$0$ to $0.292$ GeV). Fig.~(\ref{piRe_M}) and (\ref{piRe_M2}) present
the corresponding results of real part for different baryons.
Summing all the baryon loops the total self-energy is represented in
Fig.~(\ref{pi_B_kT}) where upper and lower panel are exhibiting the imaginary
and real part respectively for different set of ($\vk$, $T$ and $\mu_N$). 
Similar to mesonic loops, Im$\Pi_B$ in low mass region is sensitively
enhanced for finite values of $\vk$ and it also increases with the increasing
of $T$ and $\mu_N$. The upper panel of Fig.~(\ref{pi_B_kT}) clearly demonstrates
this fact. The non-zero values of Im$\Pi_B(M)$ for $NB$ loops 
(or Im$\Pi_M(M)$ for $\pi\Phi$ loops) in the Landau cuts, $0<M<m_B-m_N$
(or $0<M<m_\Phi-m_\pi$) basically interprets the statistical probability
of pion due to scattering with medium constituents~\cite{Weldon}. The possible scattering processes
are as follows. During propagation in the medium, pion may 
disappear by absorbing a thermalized $N$ (or $\pi$) from the medium to create a 
thermalized $B$ (or $\Phi$). Again pion may appear by absorbing 
a thermalized $B$ (or $\Phi$) from the medium as well as by emitting a 
thermalized $N$ (or $\pi$). Hence these kind of scattering processes 
provide the statistical probability to pion
and this probability in the low $M$ region is quite high 
for finite values of $\vk$, $T$ and $\mu_N$. From the upper
panel of Fig.~(\ref{pi_mes_kT}) and (\ref{pi_B_kT}), one can notice that 
the low mass probability of pion for mesonic loops is dominating
over that of the baryonic loops.

\begin{figure}
\begin{center}
\includegraphics[scale=0.35]{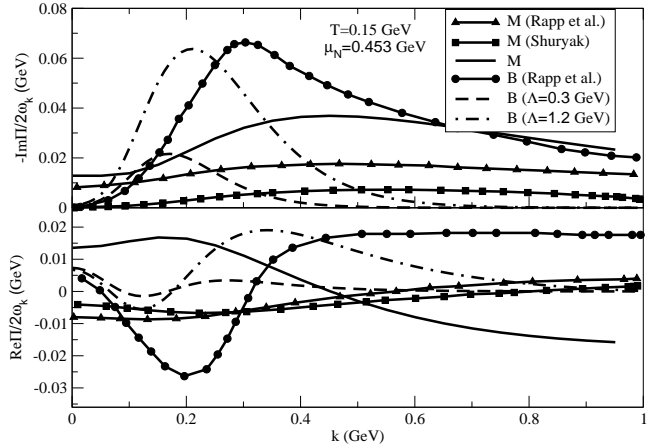}
\end{center}
\caption{The imaginary (upper panel) and real part (lower panel)
of pion potential for mesonic matter, obtained by Shuryak~\cite{Shuryak} 
(solid line with square), Rapp et al.~\cite{Rapp_M} (solid line with triangle)
and for baryonic matter, obtained by Rapp et al.~\cite{Rapp_B} (solid line with 
circle). Our corresponding results for meson loops, $N\Delta$ loop at 
$\Lambda=0.3$ and $1.2$ GeV are shown by solid, dashed 
and dash-dotted line respectively.}
\label{Self_comp}
\end{figure}

\begin{figure}
\begin{center}
\includegraphics[scale=0.35]{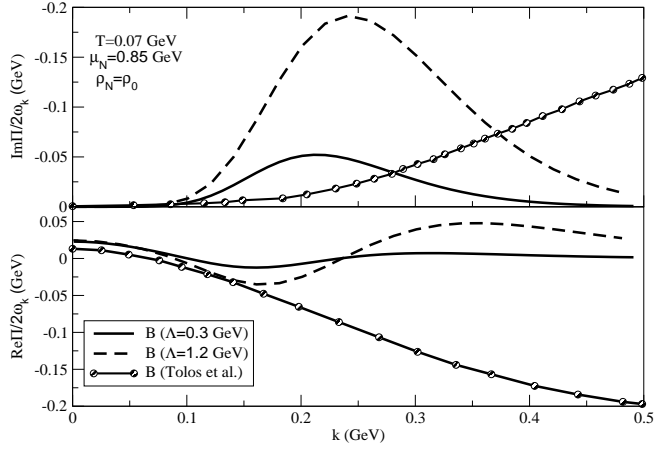}
\end{center}
\caption{The imaginary (upper panel) and real part (lower panel)
of pion potential at temperature, $T=0.07$ GeV and nucleon
density, $\rho_N=\rho_0$. Solid (for $\Lambda=0.3$ GeV) and 
dashed line (for $\Lambda=1.2$ GeV) show our results for $N\Delta$ loop
whereas the solid line with circle shows the corresponding results 
of Tolos et. al~\cite{Tolos}.}
\label{Self_comp2}
\end{figure}

To compare with some earlier results from Refs.~\cite{Shuryak,Rapp_M,Rapp_B},
we have generated the numerical values of pion optical potential 
$\frac{\Pi(k_0=\om_k,\vk)}{2\om_k}$, where $\om_k=\sqrt{\vk^2+m_\pi^2}$.
Its imaginary (upper panel) and real (lower panel) part
for mesonic (solid line) and baryonic (dotted line) matter are plotted
against $\vk$ in Fig.~(\ref{Self_comp}). For mesonic matter, our
imaginary part of pion potential is qualitatively similar with the results,
obtained by Shuryak~\cite{Shuryak} (solid line with square) and 
Rapp et al.~\cite{Rapp_M} (solid line with triangle)
but quantitatively larger than them. One of the reason for this quantative difference
may be that
our results are originated from the effective $\pi\pi\rho$ and $\pi\pi\sigma$ interactions.
In baryonic matter (considering $N$ and $\Delta$ in the medium), 
our imaginary and real part both follow approximately similar $\vk$ dependency
as followed in Ref.~\cite{Rapp_B} (solid line with circle).
In Fig~(\ref{Self_comp2}), We have also compared our 
baryonic results with the results of Tolos et al.~\cite{Tolos},
where the pion self-energy incorporates a momentum independent s-wave
and a momentum dependent p-wave piece originated from
the one and two nucleon-hole and delta-hole excitations,
plus short-range correlations.
Unlike to the results of Tolos et al.~\cite{Tolos}, our results
and the results of Rapp et al.~\cite{Rapp_B} exhibit a non-monotonic
momentum dependence of pion optical potential with a peak structure
in its imaginary part. From the Eq.~(\ref{Pi_B_bar}) this nature can
be grossly understood. The numerical integration of thermal distribution
will roughly be close to a sine hyperbolic function of momentum, which
will be a rapidly increasing function whereas the $1/16\pi\vk$ term
before the integral will make it to be a decreasing function. Hence
the net effect of these two opposite dependence of momentum create
such kind of non-monotonic momentum dependent potential.
In Ref.~\cite{Tolos}, incorporating the extra two loop 
kind of nucleon-hole and delta-hole excitations
as well as an explicit function of $\Delta$ width with external variable 
may be one of the key reason for exhibiting the monotonically increasing
behavior of pion optical potential with momentum.

\begin{figure}
\begin{center}
\includegraphics[scale=0.35]{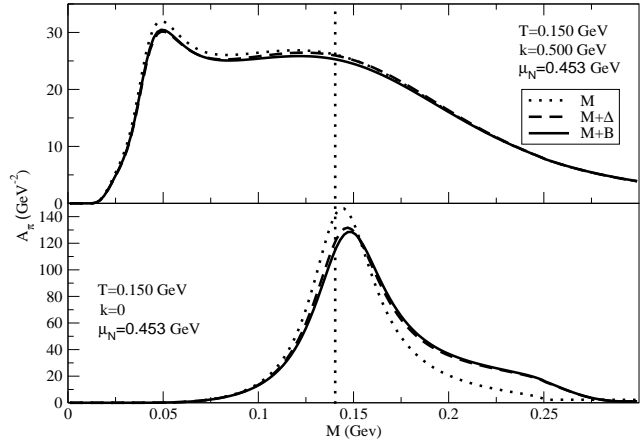}
\end{center}
\caption{The contributions in pion spectral function from
meson (dotted line), meson + $\Delta$ (dashed line) and meson + all baryon
(solid line) loops
are displayed for $(T, \mu_N, \vk)=(0.15, 0.453, 0)$ GeV (lower panel) 
and $(T, \mu_N, \vk)=(0.15, 0.453, 0.5)$ GeV (upper panel).}
\label{pi_spec1}
\end{figure}

Using the imaginary and real part of pion self-energy
in the Eq.~(\ref{pi_spec}), pion spectral function can be numerically
estimated. The contributions of
meson and baryon loops in the pion spectral function
are cumulatively shown in Fig.~(\ref{pi_spec1}).
After adding $N\Delta$ as well as the other
$NB$ loops with the meson loops, the peak of the pion spectral
function is gradually reduced and shifted toward the higher $M$.
This fact can be noticed very well in the lower panel of Fig.~(\ref{pi_spec1})
for the parameters $(T, \mu_N, \vk)=(0.15, 0.4, 0)$ GeV.
From the lower panel of Fig.~(\ref{pi_B_kT}), the positive 
values of Re$\Pi_B$ around the pion pole already indicates the fact.
By changing the three momentum $\vk$ from $0$ to $0.5$ GeV,
the peak strength of the pion spectral function has been
reduced significantly, which can be observed in the upper
panel of Fig.~(\ref{pi_spec1}). The reason for such kind of
reduction is because of the enhancement
of Im$\Pi$ at finite $\vk$, which has been observed in the 
upper panel of Fig.~(\ref{pi_mes_kT}) and (\ref{pi_B_kT}). 
This kind of softening of the pion pole for increasing momentum
also observed in some earlier works~\cite{Rapp_picloud,Tolos}. 
From Fig.~(\ref{pi_spec1}), we also observe that the $N\Delta$ loop dominates
over the rest of the $NB$ loops to determine the $\mu_N$ dependency
of pion spectral profile. The Landau cut structures of the other
$NB$ loops are dominantly contributed to the higher $M$ regions,
which are slightly away from the pion pole. Hence, their contributions
may be considered as a negligible correction part in the pion spectral function
with respect to the contribution of $N\Delta$ loop.
Similar dominance of $N\Delta$ loop over the other baryon loops 
was also noticed by Post et al.~\cite{Leupold_pi}.
However, by increasing the pion momentum $\vk$ this correction part of Im$\Pi/m_\pi$
can be dominant over the contribution of $N\Delta$ loop as shown in the Table~(2).
\begin{table}[h]
\begin{center}
\begin{tabular}{|c|c|c|c|c|}
\hline
& & & & \\
$\vk$ & Im$\Pi/m_\pi$ ($N\Delta$) & Im$\Pi/m_\pi$ ($NB$) & Re$\Pi/m_\pi$ ($N\Delta$) & Re$\Pi/m_\pi$ ($NB$) \\
& & & &  \\
\hline
& & & & \\
0 & 1.4 MeV & 1.4 MeV & 8.4 MeV & 11.4 MeV \\
& & & & \\
& & & & \\
300 MeV & 38.6 MeV & 43.5 MeV & 15.1 MeV & 15.9 MeV \\
& & & & \\
& & & & \\
500 MeV & 3.3 MeV & 10.1 MeV & 5.6 MeV & 3.4 MeV \\
& & & & \\
\hline
\end{tabular}
\caption{Table shows imaginary and real part of pion self-energy
coming from $N\Delta$ loop only and all $NB$ loops (including $N\Delta$)
for different values of pion momentum $\vk$.}
\end{center}
\label{Tab2}
\end{table}

\begin{figure}
\begin{center}
\includegraphics[scale=0.35]{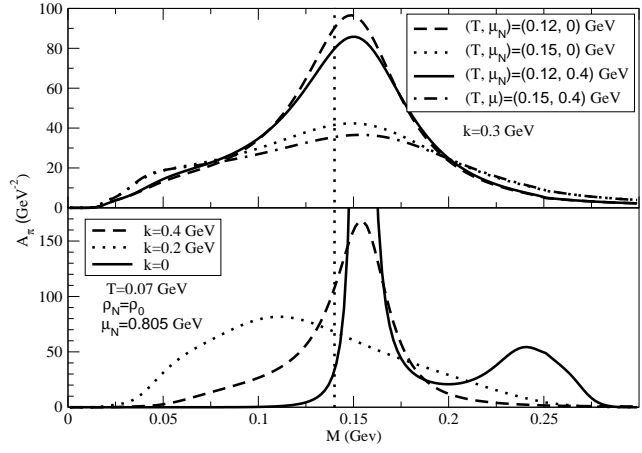}
\end{center}
\caption{Full pion spectral function for different sets of ($T, \mu_N$)
and $\vk$ are shown in upper and lower panel respectively.}
\label{pi_spec2}
\end{figure}

Using the total pion self-energy after summing all the meson
and baryon loops, we can get a full spectral function of $\pi$
meson at finite temperature and density. For different sets 
of ($T, \mu_N$) and $\vk$, this explicit structure of pion
spectral function are respectively demonstrated in upper and lower 
panel of Fig.~(\ref{pi_spec2}).
Due to noticeable Landau cut contribution of pion in the low
mass region (below its pole position), there will be some
Bose enhanced probability of $\rho\rightarrow\pi\pi$ (along with
$\pi\pi\rightarrow\rho$) below the vacuum threshold $2m_\pi$,
which may have some contribution in the low mass dilepton enhancement.
When we will apply this explicit in-medium pion spectral function
to generate dilepton spectra via $\rho$ meson modification, the probability
of pion at high momentum will definitely be suppressed due to its statistical
weight governed by the Bose-Einstein distribution function.
Hence, a selective finite momentum range will be relevant in the dilepton
spectra and the low mass enhanced probability of pion in that particular range
may have some influence in the low mass dilepton spectra. 
This  is our next interest of our future work.
%

\section{Summary and conclusion}
To summarize, the in-medium changes of pion propagation
due to different mesonic and baryonic quantum fluctuation
at finite temperature is investigated in the RTF. 
Since effective hadronic Lagrangian for $\pi\pi\sigma$ and 
$\pi\pi\rho$ interactions can alternatively describe the vacuum
$\pi\pi$ scattering cross section, hence $\pi\sigma$ and
$\pi\rho$ loops have been considered as relevant mesonic
quantum fluctuations of pion propagation in the effective theory. 
The imaginary and real
part of pion self-energy for those mesonic loops
have been evaluated in the RTF.
To account the baryonic quantum fluctuation of
pion propagation at finite temperature, an extensive
set of nucleon-baryon loops are taken. 
After summing all the mesonic and baryonic loop contributions 
the total self-energy of pion in hot and dense nuclear matter 
has been numerically estimated. 
The imaginary and real part of this total self-energy 
determine the Breit-Wigner type structure of 
in-medium pion spectral function, which is appeared to be
slightly complex in nature because of the non-trivial branch
cut structures of self-energy function. 
At high temperature as well as density, the pion spectral profile
are broadened with their attenuated peak structures.
The main punch point of this article is the low mass probability
of pion which is highly sensitive with pion's momentum ($\vk$)
and also with the medium parameters ($T$ and $\mu_N$). This
probability is completely appeared because of medium as Landau
cuts of pion self-energy for different mesonic and baryonic loops are
responsible for this contribution. 
Because of this low mass probability of pion, a non-zero
Bose enhanced probability of $\rho\rightarrow\pi\pi$ (along with
$\pi\pi\rightarrow\rho$) can be received below the vacuum threshold $2m_\pi$,
which promise to contribute in the low mass dilepton enhancement.
However, a particular range of pion momentum will be responsible for this 
low mass dilepton enhancement when we will apply this in-medium pion spectral function
on the dilepton spectra via $\rho$ meson modification. 
This efforts of calculating the low mass dilepton rate, which
will be coming from the $\rho$ meson self-energy for modified $\pi\pi$
loop, are in progress and will be addressed in due course.

{\bf Acknowledgment :} The work is financially supported
by Fundacao de Amparo a Pesquisa do Estado de Sao Paulo, 
FAPESP (Brazilian agencies), under Contract No. 2012/16766-0.
I am very grateful to Prof. Gastao Krein for his academic
and non-academic support during my postdoctoral period in Brazil.

\end{document}